\documentclass{svjour3}                     
\smartqed  
\usepackage{graphicx}
\usepackage[sort&compress,numbers]{natbib}
\journalname{Few-Body Systems (FB20)}
\begin{document}

\title{
Quark mass dependence of the ground-state octet baryons in next-to-next-to-next-to-leading order covariant baryon chiral perturbation theory
\thanks{Presented at the 20th International IUPAP Conference on Few-Body Problems in Physics, 20 - 25 August, 2012, Fukuoka, Japan}
}
\subtitle{
}


\author{X.-L. Ren \and  L. S. Geng \and  J. Martin Camalich \and J. Meng  \and H. Toki
}


\institute{X.-L. Ren \at
Research Center for Nuclear Science and Technology \& School of Physics and
Nuclear Energy Engineering, Beihang University, Beijing 100191, China\\
           \and
           L. S. Geng \at
      Research Center for Nuclear Science and Technology \& School of Physics and
Nuclear Energy Engineering, Beihang University, Beijing 100191, China\\
            \email{lisheng.geng@buaa.edu.cn}
            \and
           J. Martin Camalich \at
           Department of Physics and Astronomy, University of Sussex,
BN1 9QH, Brighton, UK
           \and
           J. Meng \at
           Research Center for Nuclear Science and Technology \& School of Physics and
Nuclear Energy Engineering, Beihang University, Beijing 100191, China \\
State Key Laboratory of Nuclear Physics and Technology, School of Physics,
Peking University, Beijing 100871, China\\
Department of Physics, University of Stellenbosch,
 Stellenbosch, South Africa
           \and
           H. Toki \at
           Research Center for Nuclear Physics (RCNP), Osaka University, Ibaraki, Osaka 567-0047, Japan
}

\date{Received: date / Accepted: date}

\maketitle

\begin{abstract}
 We report on a recent study of
  the ground-state octet baryon masses  using the covariant baryon chiral perturbation theory with
  the extended-on-mass-shell renormalization scheme up to next-to-next-to-next-to-leading order. By adjusting the available $19$ low-energy constants, a good fit of
  the $n_f=2+1$ lattice quantum chromodynamics results from the PACS-CS, LHPC, HSC, QCDSF-UKQCD and NPLQCD collaborations is achieved.
\end{abstract}

\section{Introduction}
\label{intro}
 Recently, the lowest-lying baryon spectrum, composed of up, down and strange quarks, has been studied by various lattice quantum
 chromodynamics (LQCD) collaborations~\cite{Alexandrouitetal.2009_Phys.Rev.D80_114503, Durritetal.2009_Science322_1224, Aokiitetal.(PACS-CSCollaboration)2009_Phys.Rev.D79_034503, Aokiitetal.(PACS-CSCollaboration)2010_Phys.Rev.D81_074503, Walker-Louditelal.2009_Phys.Rev.D79_054502, Linitetal.(HSCCollaboration)2009_Phys.Rev.D79_034502, Bietenholzitetal.2010_Phys.Lett.B690_436--441, Bietenholzitetal.(QCDSF-UKQCDCollaboration)2011_PhysicalReviewD84_054509, Beaneitetal.(NPLQCDCollaboration)2011_PhysicalReviewD84_014507}.
At present, most of the LQCD simulations are still performed in a finite hypercube and  with larger than physical light quarks masses~\cite{Fodor2012_Rev.Mod.Phys.84_449}, the final results can only be obtained by extrapolating to the physical point (chiral extrapolation) and infinite space-time (finite-volume corrections). Baryon chiral perturbation theory (BChPT) provides a useful framework to perform such extrapolations and to study the induced uncertainties.

In the past decades, the ground-state (g.s.) octet baryon masses have been studied extensively in BChPT~\cite{Jenkins1992_Nucl.Phys.B368_190,Bernard1993_Z.PhysikC60_111--119, Banerjee1995_Phys.Rev.D52_11, Borasoy1996_Ann.Phys.(N.Y.)254_192--232,Walker-Loud2004_Nucl.Phys.A747_476--507, Ellis1999_Phys.Rev.C61_015205, Frink2004_JHEP07_028, Frink2005_Eur.Phys.J.A.24_395, Lehnhart2004_J.Phys.G:Nucl.Part.Phys.31_89, MartinCamalich2010_Phys.Rev.D82_074504, Young2010_Phys.Rev.D81_014503, Semke2006_Nucl.Phys.A778_153--180,Semke2007_Nucl.Phys.A789_251--259, Semke2012_Phys.Rev.D85_034001,Bruns:2012eh,Lutz:2012mq}.
However, up to now, a simultaneous description of all the $n_f=2+1$ LQCD data with finite-volume effects taken into account self-consistently is still missing. Such a study is necessary for a clarification of the convergence problem and for testing the consistency between different LQCD simulations~~\cite{Beringeritetal.(ParticleDataGroup)2012_Phys.Rev.D86_010001}. Furthermore, it helps determine/constrain the many unknown LECs of BChPT at next-to-next-to-next-to-leading order (N$^3$LO).

In this work we study the g.s. octet baryon masses using the EOMS-BChPT up to N$^3$LO. Finite-volume corrections to the lattice data are calculated self-consistently~\cite{Geng:2012} and are found to be important in order to obtain a good fit of the LQCD data. Unlike Refs.~\cite{Young2010_Phys.Rev.D81_014503,Semke2012_Phys.Rev.D85_034001}, the virtual decuplet contributions are not explicitly included, because its effects can not be disentangled from those of virtual octet baryons due to the large number of unknown LECs. In order to fix all the $19$ LECs and test the consistency of current LQCD calculations, we perform a simultaneous fit of all the publicly available $n_f=2+1$ LQCD data from the PACS-CS~\cite{Aokiitetal.(PACS-CSCollaboration)2009_Phys.Rev.D79_034503}, LHPC~\cite{Walker-Louditelal.2009_Phys.Rev.D79_054502}, HSC~\cite{Linitetal.(HSCCollaboration)2009_Phys.Rev.D79_034502}, QCDSF-UKQCD~\cite{Bietenholzitetal.(QCDSF-UKQCDCollaboration)2011_PhysicalReviewD84_054509} and NPLQCD~\cite{Beaneitetal.(NPLQCDCollaboration)2011_PhysicalReviewD84_014507} collaborations.

\section{Results and disucssion}

The details of the calculation can be found in Ref.~\cite{Ren:2012aj}. Here we only briefly summarize the main results.
Up to N$^3$LO, there are 19 LECs to be determined. To make sure that N$^3$LO
BChPT is suitable for the description of the LQCD data, we have chosen two groups of LQCD data according to the following two
criteria. For the first group, we require $M^2_\pi<0.25$ GeV$^2$ and $M_\phi L>4$ with $\phi=\pi, K,\eta$ and $L$ the spacial
dimension of the LQCD simulation.
For the second, we require $M^2_\pi<0.5$ GeV$^2$, $M_K^2<0.7$ GeV$^2$ and $M_\phi L>3$. In the following, we refer to these two groups of LQCD data as
data Set-I and Set-II.

We perform a $\chi^2$ fit to the LQCD data  and the physical octet baryon masses by varying the $19$ LECs.
The so-obtained values of the LECs from the best fits are listed in Table~6 of Ref.~\cite{Ren:2012aj}.
For the sake of comparison, we have fitted Set-I using the NLO and NNLO EOMS-BChPT. The values of the LECs $b_0$, $b_D$, $b_F$, and $m_0$ are also tabulated in Table~6 of Ref.~\cite{Ren:2012aj}. An order-by-order improvement is clearly seen, with decreasing $\chi^2/\mathrm{d.o.f.}$ at each increasing chiral order. Apparently, only using the $\mathcal{O}(q^3)$ chiral expansion, one cannot describe simultaneously  the LQCD data from the five collaborations. The corresponding $\chi^2/{\rm d.o.f.}$ is about $8.6$. On the other hand, in the N$^3$LO fit of lattice data Set-I and experimental octet baryon masses, $\chi^2/{\rm d.o.f.} = 1.0$. In addition, the values of the fitted LECs (named Fit I) all look very natural. Especially, the baryon mass in the chiral limit $m_0 = 880$ MeV seems to be consistent with the SU(2)-BChPT  value \cite{Procura2003_Phys.Rev.D69_034505,Procura2006_Phys.Rev.D73_114510}. The fit to data Set-II yields a $\chi^2/{\rm d.o.f.}$ about $1.6$ and the fitted LECs look similar to those from Fit I except
$b_2$, $b_6$, $b_7$ and $b_8$. The increased $\chi^2$/d.o.f. indicates that data Set-II is a bit beyond the region of applicability of N$^3$LO BChPT.

\begin{figure*}[t]
\centering
  \includegraphics[width=10cm]{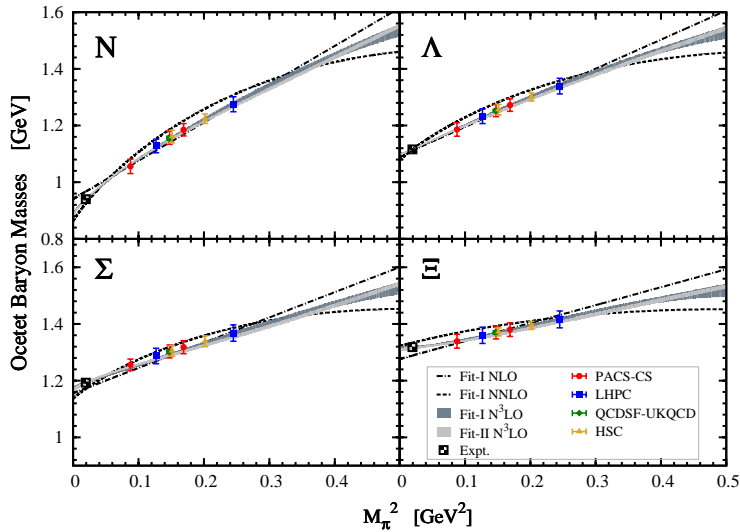}\\
  \caption{(Color online). The lowest-lying baryon octet masses as functions of the pion mass squared.
  The two  bands correspond to the best $\mathcal{O}(q^4)$ fit to lattice data Set-I and Set-II,
  and the dot-dashed lines and dashed lines are the best NLO and NNLO fits to lattice data Set-I.
    In obtaining the ChPT results, the strangeness quark mass has been set to its physical value.
    The lattice numbers are projected ones with N$^3$LO BChPT with the LECs determined from the best fit to Set-I and
    their strange quark mass is also set to the physical value. (Taken from Ref.~\cite{Ren:2012aj}.)
  }
  \label{NLO-NNLO-phy}
\end{figure*}

In Fig.~\ref{NLO-NNLO-phy}, setting the strange-quark mass to its physical value, we plot the light-quark mass evolution of $N$, $\Lambda$, $\Sigma$ and $\Xi$ as functions of $M_{\pi}^2$ using the LECs from Table~6 of Ref.~\cite{Ren:2012aj}. We can see that the NNLO fitting results are more curved and do not describe well lattice data Set-I. On the contrary the two N$^3$LO fits, named Fit I and Fit II, both can give a good description of lattice data Set I. The rather linear dependence of the lattice data on $M_\pi^2$ at large light quark masses, which are exhibited both by the lattice data~\cite{Walker-Louditelal.2009_Phys.Rev.D79_054502} and
reported by other groups, is clearly seen.

\section{Summary and Conclusions}

We have studied the lowest-lying octet baryon masses with the  EOMS BChPT  up to N$^3$LO. The unknown LECs are determined by a simultaneous fit of the latest $n_f=2+1$  LQCD simulations  from the PACS-CS, LHPC, HSC, QCDSF-UKQCD and NPLQCD collaborations. Finite-volume corrections are calculated self-consistently. It is shown that the eleven lattice data sets with  $M_{\pi}^2<0.25$ GeV$^2$ and $M_{\phi}L>4~(\phi=\pi,~K,~\eta)$ can be fitted with a $\chi^2$/d.o.f.~=~1.0.  Including more lattice data with larger pion masses or smaller volumes deteriorates the fit a bit but still yields a reasonable $\chi^2/\mathrm{d.o.f}=1.6$.

 Our studies confirm that covariant BChPT in the three flavor sector converges as expected, i.e., relatively slowly as dictated by $m_K/\Lambda_{\chi\mathrm{SB}}$ but with clear improvement order by order, at least concerning the octet baryon masses. A successful simultaneous fit of all the latest $n_f=2+1$  LQCD simulations indicates that the LQCD results are consistent with each other, though their setups are quite different.

\begin{acknowledgements}
This work was partly supported by the National
Natural Science Foundation of China under Grants No. 11005007, No. 11035007, and No. 11175002,  the Fundamental Research Funds for the Central Universities, and
the Research Fund  for the Doctoral Program of Higher Education under Grant No. 20110001110087.
JMC acknowledges the Spanish Government and FEDER funds under contract FIS2011-28853-C02-01 and the STFC [grant number ST/H004661/1] for support.
\end{acknowledgements}


\end{document}